\documentclass{amsart}
\usepackage[dvips]{graphicx}
\usepackage{amsmath}
\usepackage{amsfonts}
\usepackage{amssymb}
\begin{document}

\title[]{Self-similarity symmetry and fractal \\ distributions in iterative dynamics 
\\ of dissipative mappings}

\author[]{Vladimir ZVEREV~$^\dag$, Boris RUBINSTEIN~$^\ddag$}

\address{$^\dag$~Ural State Technical University, Ekaterinburg, 620002, Russia, E-mail: zverev@dpt.ustu.ru}

\address{$^\ddag$~Stowers Institute for Medical Research, 1000 E.50th St., Kansas City, MO, 64110, USA,
E-mail: bru@stowers-institute.org}

\maketitle

%%%%%%%%%%%%%%%%%%%%%%%%%%%%%%%%%%%%%%%%%%%%%%%%%%%%%%%%%%%%%%%%%%%%%%
\begin{abstract}
We consider transformations of deterministic and random
signals governed by simple dynamical mappings. It is shown that the
resulting signal can be a random process described in terms of
fractal distributions and fractal domain integrals. In typical cases
a steady state satisfies a dilatation equation, relating an unknown
function $f(x)$ to $f(\kappa x)$ (for example, ${f(x)=g(x)f(\kappa
x)}$). We discuss simple linear models as well as nonlinear systems
with chaotic behavior including dissipative circuits with delayed
feedback. \\
\textit{2000 Mathematical Subject Classification} 37F25; 28A80; 37H50; 70K55; 37D45 \\
\textit{Keywords} Scaling; Fractals; Stochastic difference equations; Nonlinear difference equations; Chaotic behavior 
\end{abstract}
%%%%%%%%%%%%%%%%%%%%%%%%%%%%%%%%%%%%%%%%%%%%%%%%%%%%%%%%%%%%%%%%%%%%%%

\section{Introduction}

One of the most interesting and significant questions in nonlinear
dynamics is the problem of noise influence on chaotic motion.
Traditional deterministic models provide a fair description of
averaged motion only for stable trajectory systems. In case of
unstable motion random fluctuations may grow, so that non-stochastic
descriptions fail which asks for deeper understanding and
investigation of noisy dynamics.

In this work we consider an evolution of simple linear and nonlinear
noisy discrete dynamical systems. We concentrate on the important
peculiarities of the stochastic evolution: \textit{self-similarity
symmetry} and \textit{fractal distribution generation}.

Consider a stochastic mapping:
\begin{equation}
X_{N+1} = \xi_N^{\mbox{\scriptsize{fl}}} + F(X_N)
\label{eq1}
\end{equation}
\noindent
in real $d$-dimensional space. Assuming $\xi_N^{\mbox{\scriptsize{fl}}}$ being a sequence of
uncorrelated random variables, we can write the
Kolmogorov-Chapman equation, which for large $N\to\infty$ reduces to
the equation for the stationary (asymptotic) distribution
$P_{\mbox{\scriptsize{st}}}$ that can be written in the form:
\begin{equation}
P_{\mbox{\scriptsize{st}}}(X) = \iint dY \, dZ \,
P_{\mbox{\scriptsize{fl}}}(X-Y) \, \delta(Y-F(Z)) \,
P_{\mbox{\scriptsize{st}}}(Z).
\label{eq2}
\end{equation}
\noindent
The corresponding equation for the distribution function Fourier transform
\begin{equation*}
\Psi_{\mbox{\scriptsize{st}}} (U) = \int dX P_{\mbox{\scriptsize{st}}}
(X) \exp \{ - i \langle X,U \rangle \}
%\label{eq3}
\end{equation*}
\noindent
is given by
\begin{gather}
\Psi_{\mbox{\scriptsize{st}}} (U) =\Psi_{\mbox{\scriptsize{fl}}}(U) \int dV  \,
\sigma(U,V) \, \Psi_{\mbox{\scriptsize{st}}} (V),
\label{eq4} \\
\sigma(U,V) = (2 \pi)^{-d} \int dY \exp \{i \langle Y,V \rangle - i \langle F(Y), U \rangle \},
\label{eq5}
\end{gather}
\noindent
where the scalar product takes the form: $\langle X,U \rangle = x_1 u_1+ \cdots +x_d u_d $.

\section{Easy cases: linear deterministic and noisy mappings}

We can consider a mapping \eqref{eq1} as a transformation of a
$d$-dimensional signal in a circuit with dissipation. It means that
the mapping is contractive and its Jacobian satisfies the condition
$| \det \{\partial F_i / \partial x_j \}| < 1$. A steady state
solution corresponds to the stationary regime of deterministic or
stochastic motion. As a preliminary we briefly discuss several
simple cases.

$\bullet \,$\textbf{Linear deterministic dissipative mapping.} Let
us suppose that a signal $X$ subject to a dissipative contraction:
$X \to \kappa X$, $\kappa < 1$, and mixed with an external signal:
$X \to A+X$. In this case we have
\begin{gather}
F(X)=A+\kappa X,  \qquad 0<\kappa<1,
%\label{eq6}
\notag \\
P_{\mbox{\scriptsize{fl}}} (X)= \delta (X),
\notag
%\label{eq7}
\end{gather}
and Eqs.\eqref{eq2}, \eqref{eq4} take the form of \textit{dilatation (scaling) equations}:
\begin{gather}
P_{\mbox{\scriptsize{st}}} (X) = \kappa^{-d} P_{\mbox{\scriptsize{st}}}
\left(\kappa^{-1} (X-A) \right),
%\label{eq8}
\notag \\
\Psi_{\mbox{\scriptsize{st}}} (U) = \exp \{- i \langle A,U \rangle \} \,
\Psi_{\mbox{\scriptsize{st}}} (\kappa U).
%\label{eq9}
\notag
\end{gather}
Utilizing a normalization $\Psi_{\mbox{\scriptsize{st}}} (0) = 1$
and using an iteration procedure we find the solution:
\begin{gather}
\Psi_{\mbox{\scriptsize{st}}} (U) = \prod \limits_{\gamma=0}^{\infty} \exp
\left\{ - i \kappa^\gamma \langle A,U \rangle \right\} =  \exp \Bigl\{ - i \langle A, U \rangle
\sum \limits_{\gamma=0}^{\infty} \kappa^\gamma \Bigr\} =
\exp \left\{ - i \frac {\langle A, U \rangle}{1-\kappa} \right\},
%\label{eq10}
\notag \\
P_{\mbox{\scriptsize{st}}} (X) = \delta \left( X - A/(1 -
\kappa) \right),
%\label{eq11}
\notag
\end{gather}
\noindent
which describes coherent accumulation in a circuit.

$\bullet \,$\textbf{Linear stochastic dissipative mapping with a
Gaussian noise term.} Assuming
\begin{gather}
F(X)=A+\kappa X,
%\label{eq12}
\notag \\
P_{\mbox{\scriptsize{fl}}} (X) = P_{\mbox{\scriptsize{g}}}
(X,R)= (\pi R)^{-d/2} \, \exp \left\{ - {\langle X,X \rangle}/R \right\},
%\label{eq13}
\notag
\end{gather}
\noindent
and solving the equations
\begin{gather}
P_{\mbox{\scriptsize{st}}} (X) = \kappa^{-d} \int dY P_{\mbox{\scriptsize{g}}}
(X-Y,R) P_{\mbox{\scriptsize{st}}} \left( \kappa^{-1} (Y-A) \right),
%\label{eq14}
\notag \\
\Psi_{\mbox{\scriptsize{st}}} (U) =  \exp \{ - \tfrac{1}{4} R \langle U,U \rangle -
i \langle A,U \rangle \} \, \Psi_{\mbox{\scriptsize{st}}} (\kappa U)
%\label{eq15}
\notag
\end{gather}
using the same procedure we obtain the solution
\begin{gather}
\Psi_{\mbox{\scriptsize{st}}} (U) = \prod \limits_{\gamma=0}^{\infty} \exp
\left\{ - i \kappa^\gamma \langle A,U \rangle - \frac{1}{4} \, R \, \kappa^{2 \gamma} \,
\langle U,U \rangle \right\} =  \exp \left\{ - i \, \frac {\langle A, U \rangle}{1-\kappa} -
\frac{1}{4} \, R \, \frac {\langle U, U \rangle}{1-{\kappa^2}} \right\},
%\label{eq16}
\notag \\
P_{\mbox{\scriptsize{st}}} (X) = P_{\mbox{\scriptsize{g}}}
\left( X - A/(1 - \kappa), \, R/(1 - {\kappa}^2) \right).
%\label{eq17}
\notag
\end{gather}
\noindent It can be viewed as a result of two independent additive
processes -- for both coherent and random components of the signal.

$\bullet \,$\textbf{Linear stochastic dissipative mapping with
Gaussian and Kubo-Andersen stochastic terms.} Let us consider a model
with a mixed stochastic term: $\xi_N^{\mbox{\scriptsize{fl}}} = \xi_N^{\mbox{\scriptsize{g}}} +
\xi_N^{\mbox{\scriptsize{KA}}}$. In this case:
\begin{gather}
F(X)=\kappa X,
%\label{eq18}
\notag \\
P_{\mbox{\scriptsize{fl}}} = P_{\mbox{\scriptsize{g}}} \ast P_{\mbox{\scriptsize{KA}}},
\qquad P_{\mbox{\scriptsize{KA}}} (X)= \sum \limits_{k=0}^{L-1} p_k \delta(X-X_k),
%\label{eq19}
\notag
\end{gather}
where $ \ast $ denotes the convolution. Now
Eqs.\eqref{eq2},\eqref{eq4} take the form:
\begin{gather}
P_{\mbox{\scriptsize{st}}} (X) = \frac{1}{\kappa^d} \int dY P_{\mbox{\scriptsize{g}}}
(X-Y,R)  \, \sum \limits_{k=0}^{L-1} p_k P_{\mbox{\scriptsize{st}}} \left( \frac{Y-X_k}{\kappa} \right),
\label{eq20} \\
\Psi_{\mbox{\scriptsize{st}}} (U) =  \exp \{ - \frac{1}{4} R \langle U,U \rangle \} \,
\Bigl\{ \sum \limits_{k=0}^{L-1} p_k \exp (-i \, \langle X_k,U \rangle) \Bigr\} \,
\Psi_{\mbox{\scriptsize{st}}} (\kappa U).
\label{eq21}
\end{gather}
The solution of Eq.\eqref{eq21} by the iteration procedure leads to a distribution with
fractal properties:
\begin{equation}
\Psi_{\mbox{\scriptsize{st}}} (U) =  \exp \Bigl\{ - \frac{1}{4} \, R
\, \frac{\langle U,U \rangle} {1-\kappa^2} \Bigr\} \, \prod
\limits_{\gamma=0}^{\infty} \Bigl\{ \sum \limits_{k=0}^{L-1} p_k
\exp (-i \, \kappa^\gamma \, \langle X_k,U \rangle) \Bigr\}
\label{eq22}
\end{equation}
It is convenient to represent this solution  through integral over
the fractal domain, or \textit{multifractal integral} (MFI). The MFI
concept is discussed in the next section.

\begin{figure}[tb]
\begin{center}
\includegraphics[scale=1.1]{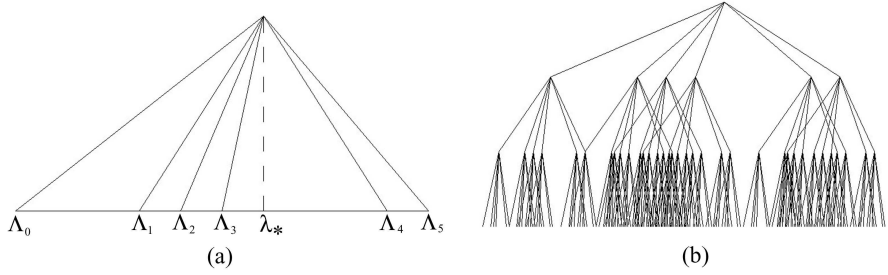}
\end{center}
\caption{ (a) The scheme of branching for $L=6$. (b) The tree for the third generation of a pre-fractal.}
\end{figure}

\section{Multifractal integrals}

Consider the fractals with $L$-branch multiplication (the case of
dichotomous branching was considered in \cite{Zverev1}). Having
assigned the values
\begin{equation*}
\lambda_{\ast} \in (0,1), \qquad 0 \equiv \Lambda_0<\Lambda_1< \cdots \Lambda_{L-1} \equiv 1
%\label{eq23}
\end{equation*}
and the contraction ratio $\kappa < 1$, we choose the law
of branching, shown in Figure 1, and define the multifractal integral of
$f(x)$, with $x \in[0,1]$, as
\begin{equation}
\int \limits_{\mathcal{L}} f(x)d \mu (x | \kappa ,\Theta ) = \lim
\limits_{n \to \infty } \sigma_n, \qquad \sigma _n = L^{- n} \sum \limits_s \Theta_n^{[s]}
f(\lambda_n^{[s]}),
\label{eq24}
\end{equation}
\noindent (existence and equality conditions for the limits in
\eqref{eq24} are formulated below). The summation in \eqref{eq24} is
made over all possible signature vectors $\mathbf{s}_n = (s_n^{(1)},
\ldots, s_n^{(n)} )$ with components $ s_n^{(i)} \in \{
\bar{\Lambda}_0, \bar{\Lambda}_1, \ldots, \bar{\Lambda}_{L-1} \}$ ,
where $\bar{\Lambda}_k = \Lambda_k - \lambda_{\ast}$. The arguments
of the function $f$ are defined as follows
\begin{equation}
\lambda_n^{[s]} = \lambda_{\ast} + (1 - \kappa )(\mathbf{s}_n ,
\mathbf{h}_n (\kappa )),
\label{eq25}
\end{equation}
\noindent where ${\rm{\bf h}}_n (\kappa) = (1, \kappa ,\kappa^2,
\ldots ,\kappa^{n - 1})$  and $({\rm{\bf a}},{\rm{\bf b}}) = a_1
b_1+ \cdot \cdot \cdot +a_n b_n$, and the quantities $\Theta
_n^{[s]}$ are functions of $\mathbf{s}_n$. Choosing normalization
condition $\sigma _n = 1$ for $f \equiv 1$ we find that
\begin{equation}
\sum \limits_{k=0}^{L-1} \Theta _n^{[{s}^k]} = L \, \Theta _{n - 1}^{[s]}
\label{eq26}
\end{equation}
\noindent for $\mathbf{s}_n^k = \mathbf{s}_{n - 1} \oplus
(\bar{\Lambda}_k)$ (where the $\oplus$ denotes a "concatenation" of
vectors). In this section we assume that the conditions
 \begin{gather}
\Theta _n^{[s]} = \prod \limits_{i = 1}^n \psi (s_n^{(i)}), \qquad
\sum \limits_{k=0}^{L-1} \psi (\bar{\Lambda}_k) = L,
\label{eq27} \\
\mbox{Im} \, \psi (\bar{\Lambda}_k) = 0, \qquad \mbox{Re} \,
\psi(\bar{\Lambda}_k) > 0, \label{eq28}
\end{gather}
\noindent hold. In this case one can interpret $p_k=
\psi(\bar{\Lambda}_k)/L$ as probabilities; the corresponding
\textit{multifractal measures} induced by the \textit{multiplicative
Besikovitch process} were considered in \cite{Feder}. Setting $X_i =
(1-\kappa) (\lambda_{\ast} + \bar{\Lambda}_i)$, we represent the
solution of Eq.\eqref{eq20} (the Fourier transform of \eqref{eq22})
through the MFI
\begin{equation*}
P_{\mbox{\scriptsize{st}}} (X) = \int \limits_{\mathcal{L}} P_{\mbox{\scriptsize{g}}}
(X-x) d \mu (x | \kappa ,\Theta ).
\notag
\end{equation*}

The following existence conditions hold for MFI:

\textbf{Theorem 1.} \textit{If a function $f(x)$, $x \in[0,1]$, is continuous
and conditions \eqref{eq27}, \eqref{eq28} hold, the sequences of the
quantities $\sigma_n$ in \eqref{eq24} converge to a limit equal to MFI.}

\textbf{Proof.} For certain $n$, the summation in \eqref{eq24} goes over the points 
of the $n$-th generation pre-fractal, which is the set of $L^n$
points with coordinates $\lambda_n^{[s]} \in (0,1)$. By arranging
the pre-fractals on parallel lines and connecting the points
according to the rule $\mathbf{s}_n \Rightarrow \mathbf{s}_{n+1}^k =
\mathbf{s}_n \oplus (\bar{\Lambda}_k)$, $k=0, \ldots L-1$, we get a
graph that represents the formation of the Cantor set for $n \to
\infty$. Let the graph contains pre-fractals of first $n+k$
generations. One can see that every point of the $n$-th generation
pre-fractal is a parent vertex for a cluster containing a tree of
the subsequent generations. The corresponding rule of branching is
given by
\begin{gather}
\lambda_{n+k}^{[s \oplus r]} = \lambda_{n}^{[s]} + \Delta_{n,k}^{[r]}, \qquad
\Delta_{n,k}^{[r]} = (1-\kappa) \kappa^n \left( \mathbf{r}_k , \mathbf{h}(\kappa) \right);
%\label{eq29}
\notag \\
- \kappa^n \, \lambda_{\ast} < \Delta_{n,k}^{[r]} < \kappa^n \, (1-\lambda_{\ast}).
%\label{eq30}
\notag
\end{gather}
As $k \to \infty$, every "small" cluster converges to a Cantor fractal, which is similar
to the full fractal.

Using definition \eqref{eq24} and conditions \eqref{eq27}, we obtain
\begin{gather}
|\sigma_{n+k}-\sigma_{n+k'}| \le |\sigma_{n+k} - \sigma_n| + |\sigma_{n+k'} -\sigma_n|,
\label{eq31} \\
\sigma_{n+k} - \sigma_n = \sum \limits_s L^{-n} \Theta_n^{[s]} \sum \limits_r L^{-k}
\Theta_k^{[r]} \left( f(\lambda _n^{[s]} + \Delta_{n,k}^{[r]})-f(\lambda _n^{[s]}) \right).
\label{eq32}
\end{gather}
Taking into account that $|\Theta_n^{[s]}| \le 1$, we get an
estimate
\begin{equation}
|\sigma_{n+k} - \sigma_n| \le \sum \limits_s L^{-n} \Theta_n^{[s]} \sum \limits_r L^{-k}
\Theta_k^{[r]} \left| f(\lambda _n^{[s]} + \Delta_{n,k}^{[r]})-f(\lambda _n^{[s]}) \right|.
\label{eq33}
\end{equation}
Let $f(x)$ be continuous on the closed interval $[0,1]$ (thus, uniformly continuous):
\begin{equation*}
\forall x \in [0,1] \quad \forall \epsilon>0 \quad \exists \delta_\epsilon \quad (\forall
\alpha: \, |\alpha| < \delta_\epsilon): \quad |f(x+\alpha)-f(x)| < \epsilon/2.
\end{equation*}
Using the condition
\begin{equation*}
\left[ \lambda_n^{[s]}- \kappa^n \lambda_{\ast}, \, \lambda_n^{[s]} + \kappa^n
(1-\lambda_{\ast}) \right] \subset \left[ \lambda_n^{[s]}- \delta_{\epsilon}, \,
\lambda_n^{[s]} + \delta_{\epsilon} \right],
%\label{eq34}
\end{equation*}
we can select $N_\epsilon$ for every $\delta_\epsilon$. Employing
the relations \eqref{eq27} and \eqref{eq31}-\eqref{eq33}, we obtain:
\begin{equation*}
\forall \epsilon>0 \quad \exists N_{\epsilon} \quad \forall
n>N_{\epsilon}: \quad |\sigma_{n+k}-\sigma_{n+k'}| < \epsilon,
%\label{eq35}
\end{equation*}
showing that $\{ \sigma_n \}$ is a Cauchy sequence and thus
converges. 
$\blacksquare$

We find an alternative representation for the MFI using the singular
distributions approach. Introducing singular functions
\begin{equation}
\mathfrak{M}_n (x | \kappa ,\Theta ) = L^{-n} \sum \limits_s \Theta_n^{[s]} \,
\delta (x - \lambda_n^{[s]}), \qquad \mathfrak{M} (x | \kappa ,\Theta ) =
\lim \limits_{n \to \infty} \mathfrak{M}_n (x | \kappa ,\Theta ),
\label{eq36}
\end{equation}
\noindent
we have:
\begin{equation*}
\int \limits_{\mathcal{L}} f(x)d \mu (x | \kappa ,\Theta ) = \int \limits_0^1 f(x) \,
\mathfrak{M} (x | \kappa ,\Theta ) dx \equiv \lim \limits_{n \to \infty} \int \limits_0^1
f(x) \, \mathfrak{M}_n (x | \kappa ,\Theta ) dx.
%\label{eq37}
\end{equation*}
\noindent
The Fourier transform of the limit in \eqref{eq36} is a solution of a \textit{dilatation equation}
\begin{gather}
\hat{\mathfrak{M}} (\omega | \kappa ,\Theta ) = \hat{\mathfrak{N}}
(\omega | \kappa ,\Theta ) \, \hat{\mathfrak{M}} (\kappa \omega | \kappa ,\Theta ),
%\label{eq38}
\notag \\
\hat{\mathfrak{N}} (\omega | \kappa ,\Theta ) = L^{-1} \, \sum \limits_{k=0}^{L-1}
\Psi(\bar{\Lambda}_k) \exp \left\{ -i \omega (1-\kappa) (\lambda_{\ast} + \bar{\Lambda}_k)\right\}
%\label{eq39}
\notag
\end{gather}
\noindent
and is equal to
\begin{equation*}
\hat{\mathfrak{M}} (\omega | \kappa ,\Theta ) = e^{-i \omega \lambda_{\ast}} \prod
\limits_{\gamma=0}^{\infty} \left[ L^{-1} \, \sum \limits_{k=0}^{L-1} \Psi(\bar{\Lambda}_k)
\exp \left\{ -i \omega (1-\kappa) \kappa^{\gamma} \,  \bar{\Lambda}_k\right\} \right].
%\label{eq40}
\end{equation*}
\noindent
Accordingly, the \textit{dilatation equation} for the
limit function \eqref{eq36} reads:
\begin{gather}
\mathfrak{M} (x | \kappa ,\Theta ) = {\kappa}^{-1} \int \limits_{-\infty}^{\infty} \mathfrak{N}
(x-y | \kappa ,\Theta ) \, \mathfrak{M} \left( \kappa^{-1} y | \kappa ,\Theta \right)  dy,
%\label{eq41}
\notag \\
\mathfrak{N} (x | \kappa ,\Theta ) = L^{-1} \, \sum \limits_{k=0}^{L-1} \Psi (\bar{\Lambda}_k)
 \, \delta \left( x-(1-\kappa) (\lambda_{\ast} + \bar{\Lambda}_k ) \right).
%\label{eq42}
\notag
\end{gather}

It is possible to define the MFI as a limit of a sequence of definite integrals. Such
definition is more illustrative but in fact is equivalent to \eqref{eq24}-\eqref{eq28}.
This approach was developed in \cite{Zverev1} for $L=2$. Here we present this formalism briefly.

Define the MFI as
\begin{equation}
\int \limits_{\mathcal{L}} f(x)d \mu (x | \kappa ,\Theta ) = \lim \limits_{n \to \infty }
\sigma_n^{\#}, \qquad \sigma _n^{\#} = 2^{- n} \sum \limits_s \Theta_n^{[s]} \prec \!\! f 
\left( \lambda _n^{[s]} \right) \!\!\succ \!\!
{\vphantom{()}}_{\kappa ^n}\, ,
\label{eq43}
\end{equation}
\noindent
where
\begin{equation*}
\prec \!\!  {f(x)} \!\!\succ \!\! {\vphantom{()}}_{\delta} = {\delta}^{-1}
\int_{x - \delta / 2}^{x + \delta / 2} f(t) dt
%\label{eq44}
\end{equation*}
\noindent and $\lambda_n^{[s]}$ is defined by \eqref{eq25} with
$L=2$, $\lambda_{\ast} = \tfrac{1}{2}$, $\Lambda_0=0$,
$\Lambda_1=1$. For $\kappa < \tfrac{1}{2}$, $\sigma_n^{\#}$ is
computed by integration over a system of nonintersecting segments
$\mathcal{L}_n = \bigcup_s l_n^{[s]} \subset [0,1]$, where
$l_n^{[s]} = [ \, \lambda_n^{[s]} - \tfrac{1}{2}
\kappa^n,\lambda_n^{[s]} + \tfrac{1}{2} \kappa ^n \, ]$. The set
$\mathcal{L}_n$ is obtained from the unit segment by the standard
Cantor procedure repeated $n$ times: one first removes from the unit
segment a middle part of length $1-2\kappa$, then the same fraction
is removed from the middle of each of the remaining segments, and so
forth. The measure is constructed by assigning to each segment
$l_n^{[s]}$ a weight factor $\Theta_n^{[s]}$ appearing in the
summation in \eqref{eq43}.

The helpful intuitive picture of the pre-fractal domain $\mathcal{L}_n$
and the cluster set $\mathcal{L} = \lim_{n \to \infty}
\mathcal{L}_n$ embedded in $[0,1]$ is no longer valid for $\tfrac{1}{2} <
\kappa < 1$, since the segments $l_n^{[s]}$ intersect in that case.
This difficulty may be resolved by replacing the segments with
rectangles and "spreading" them along the second coordinate. Define
two-dimensional MFI of $f(x,y)$, $x \in [0,1]$, $y \in [0,1]$ as
follows
\begin{equation}
\iint \limits_\mathcal{D} f (x,y) d \mu (x,y | \kappa_x, \kappa_y, \Theta) =
\lim \limits_{n \to
\infty} 2^{-n} \sum \limits_s \Theta_n^{[s]} \left\langle f \left( \lambda_{xn}^{[s]},
\lambda_{yn}^{[s]} \right) \right\rangle_{\kappa_x^n ,\kappa _y^n}.
\label{eq45}
\end{equation}
\noindent
We assume that $\lambda_{\alpha n}^{[s]}$ is determined by
the expression that follows from \eqref{eq25} by replacing $\kappa$
with $\kappa_\alpha$, $\alpha = x,y$, and that the average is taken
over the region $d_n^{[s]} = l_{xn}^{[s]} \otimes l_{yn}^{[s]}$,
$l_{\alpha n}^{[s]} = [ \, \lambda_{\alpha n}^{[s]} - \tfrac{1}{2} \kappa_\alpha
^n , \lambda_{\alpha n}^{[s]} + \tfrac{1}{2} \kappa_\alpha^n \, ]$. The
rectangular cells $d_n^{[s]}$ do not overlap if at least one of the
contraction coefficients $\kappa_\alpha$, $\alpha = x,y$, is not
larger than $\tfrac{1}{2}$. This condition is satisfied, in particular, for
$\kappa_x = \kappa$, $0 < \kappa < 1$, and $\kappa_y = \tfrac{1}{2}$, and
then
\begin{equation*}
\iint \limits_\mathcal{D} f (x) d \mu (x, y | \kappa, \tfrac{1}{2}, \Theta ) =
\int \limits_\mathcal{L} f (x) d \mu ( x | \kappa, \Theta ).
%\label{eq46}
\end{equation*}
\noindent If $\kappa_\alpha \le \tfrac{1}{2}$ for $\alpha = x$ (for
$\alpha = y$), the projection of $\mathcal{D} = \lim_{n \to \infty }
\bigcup_s d_n^{[s]} $ onto the $x$ (the $y$) axis is a standard
Cantor set that has the similarity dimension $D_\alpha = - \left(
\log_2 \kappa_\alpha \right)^{-1}$. As it was shown in \cite{Feder},
the Hausdorff-Besicovitch dimension $D$ of the whole set
$\mathcal{D}$ is equal to $D = 2 / (1 - \log_2 \kappa)$. Note that
for $\kappa = \tfrac{1}{2}$ we have $D = 1$; in this case
$\mathcal{D}$ is a diagonal of the unit square, and \eqref{eq45}
reduces to a definite integral on the diagonal.

\section{The nonlinear Ikeda mapping}

Consider a circuit with a \textit{nonlinear element} and a
\textit{delayed feedback} \cite{Zverev1, Ikeda, Singh, ZverRub1, ZverRub2, ZverRub3, Zverev2}.
Assuming $\xi^{\mbox{\scriptsize{fl}}} (t)$ being a random process with zero mean value
$ \langle\!\langle \xi^{\mbox{\scriptsize{fl}}} (t)
\rangle\!\rangle =0$, we can write a \textit{stochastic difference
equation}
\begin{equation}
X(t) = F(X(t - T_{\mbox{\scriptsize{del}}} )) + \xi^{\mbox{\scriptsize{fl}}}
(t - T_{\mbox{\scriptsize{del}}} ),
\label{eq47}
\end{equation}
\noindent
where $T_{\mbox{\scriptsize{del}}}$ is the delay time. For $X(t)$ being a two-dimensional
(complex) signal, the equation \eqref{eq47} can be rewritten in the form of mapping
\eqref{eq1}, where $X_N=X(t_0+N T_{\mbox{\scriptsize{del}}} )$,
$\xi_N^{\mbox{\scriptsize{fl}}} = \xi ( t_0 + (N+1)
T_{\mbox{\scriptsize{del}}} )$, $0 \le t_0 < T_{\mbox{
\scriptsize{del}}} $ (in this case $\langle X,U \rangle = \mbox{Re }(XU^{*})$). Suppose
that the external noise is the Ornstein-Uhlenbeck random process with small correlation
time $\tau_{\mbox{\scriptsize{cor}}} \ll T_{\mbox{\scriptsize{del}}}$ (the "rapid" Gaussian noise).
 Assuming the transformation of a slowly varying complex valued amplitude $X(t)$ as a phase shift
 $\phi \to \phi + \lambda {|X|}^2 + \theta_0$ and the dissipative
contraction $|X| \to \kappa |X|$, $\kappa < 1$, we find the
nonlinear term in \eqref{eq1}, \eqref{eq47} in the form
\begin{equation}
F(X) = 1 + \kappa X \exp (i \lambda \left| X \right|^2 + i \theta_0 )
\label{eq48}
\end{equation}
\noindent
(the \textit{Ikeda mapping} \cite{Ikeda, Singh}). As shown in \cite{ZverRub1, ZverRub2, ZverRub3},
the random phase approximation can be used in the \textit{intence phase mixing limit}
$\lambda \gg 1$, which allows to replace the
delta-function in \eqref{eq2} by
\begin{equation*}
\delta ^{(2)}(Y - F(Z)) \longrightarrow \langle\!\langle \, \delta ^{(2)}(Y - 1 -\kappa
Z e^{i \phi} ) \, \rangle\!\rangle_{\phi},
%\label{eq49}
\end{equation*}
\noindent
where $\langle\!\langle \cdots \rangle\!\rangle_{\phi}$ denotes an average over $\phi$. Then
\begin{equation*}
P_{\mbox{\scriptsize{st}}} (X) = \int dY \, P_{\mbox{\scriptsize{g}}}
\left( X - Y,R/(1 - \kappa^2) \right) P_{\mbox{\scriptsize{ch}}} (Y).
%\label{eq50}
\end{equation*}
\noindent
The radially symmetric distribution
$P_{\mbox{\scriptsize{ch}}}$ describing noise caused by dynamic
chaotization \cite{ZverRub1, ZverRub2, ZverRub3} can be written in the form
\begin{gather}
P_{\mbox{\scriptsize{ch}}} (X) = (2 \pi \kappa ^2)^{-1}
\int_0^{\infty} \beta d \beta \,\, \Xi ( \beta ) \, J_0 \left(
\beta \kappa^{-1} \left| X - 1 \right| \right)
\label{eq51} \\
\Xi ( \beta ) = J_0 ( \beta ) \Xi ( \kappa \beta ), \qquad \qquad
\Xi ( \beta ) = \prod \limits_{k=0}^{\infty} J_0 ( \beta \kappa^k ).
\label{eq52}
\end{gather}
\noindent
The integrand in \eqref{eq51} contains a factor $\Xi(\beta
)$ which is a solution of the \textit{dilatation equation}
\eqref{eq52}. In order to simplify \eqref{eq51}, \eqref{eq52}, we
approximate the Bessel function by a function with separated
"oscillatory" and monotonic parts $J_0 (x) \approx \chi (x) \left[
\sqrt{2} \cos \left( x - \tfrac{1}{4} \pi \right) \right]$. In this
approximation
\begin{gather}
f_{\mbox{\scriptsize{ch}}} (Q) \approx \frac{1}{\sqrt{2}}
\left\{ e^{-i \pi / 4} \int \limits_{\mathcal{L}} \hat{\Phi}
\! \left( \! \sqrt{Q} + \frac{2x-1}{1-\kappa}, \, Q \! \right)
\, d \mu ( x | \kappa, \Theta ) + \mbox{c.c.} \right\},
%\label{eq53}
\notag \\
\Phi (\beta ,Q) =  \frac{1}{2} \beta \, \chi \! \left( \beta \sqrt{Q} \right)
\prod \limits_{k =0}^{\infty} \chi \! \left( \beta \kappa^k \right), \qquad
\Theta_n^{[s]} = 2^{n/2} \exp \left[ - \tfrac{1}{4} i \pi ({\rm {\bf s}}_n ,
{\rm {\bf h}}_n (1)) \right],
%\label{eq54}
\notag
\end{gather}
\noindent where $\hat{\Phi}$ is the Fourier transform w.r.t. first
argument of the monotonic function $\Phi$. Notice that quantities
$\Theta_n^{[s]}$ fail to satisfy the conditions \eqref{eq28}. Thus
the Theorem 1 does not hold and we have to look for an
alternative condition of convergence.

\section{Extended class of multifractal integrals}

Let us consider the definition \eqref{eq24} of the MFI without the condition \eqref{eq28}.
Now we assume that $\Psi(\bar{\Lambda}_k)$ are complex-valued.

\textbf{Theorem 2.} \textit{Let the function $f(x)$, $x \in [0,1]$, be
differentiable arbitrary many times, also let $\exists M \,\,\forall
x \,\,\forall l \in \mbox{\textbf{N}}: \,\,|f^{(l)}(x)| < M $, and
let the conditions \eqref{eq27} hold. Then the sequences $\sigma_n$
in \eqref{eq24} converge to a limit equal to MFI.}

\textbf{Proof.} Taking into account the formula \eqref{eq25} and introducing the power expansion
\begin{equation*}
f(\lambda_n^{[s]})=\sum \limits_{l=0}^{\infty} \frac{1}{l!} f^{(l)} (\lambda_{\ast})
(1-\kappa)^l \left( \mathbf{s}_n, \, \mathbf{h}_n (\kappa) \right)^l,
%\label{eq55}
\end{equation*}
we obtain
\begin{equation}
|\sigma_n| \le \sum \limits_{l=0}^{\infty} \frac{1}{l!} | f^{(l)} (\lambda_{\ast}) | \,
(1-\kappa)^l \, \Bigl| L^{-n} \sum \limits_{s} \Theta_n^{[s]} \left( \mathbf{s}_n, \,
\mathbf{h}_n (\kappa) \right)^l \Bigr|.
\label{eq56}
\end{equation}
The use of  the generating function for the powers of the scalar product $\exp\{\xi(\mathbf{s}_n,
\, \mathbf{h}_n(\kappa))\}$ and the Leibnitz formula allows us to perform the transformations:
\begin{gather}
L^{-n} \sum \limits_{s} \Theta_n^{[s]} \left( \mathbf{s}_n, \, \mathbf{h}_n (\kappa) \right)^l =
\Bigl. \frac{d^l}{d \xi^l} \Bigl( L^{-n} \sum \limits_{s} \Theta_n^{[s]} \,\, e^{ \, \xi \,
( \mathbf{s}_n, \, \mathbf{h}_n (\kappa) )} \Bigr) \Bigr|_{\xi=0}
%\label{eq57}
\notag \\
= \Bigl. \frac{d^l}{d \xi^l} \Bigl( \prod \limits_{j=1}^{n} \Bigl[ L^{-1} \sum
\limits_{k=0}^{L-1} \Psi(\bar{\Lambda}_k) \,  e^{\, \xi \, \kappa^{j-1} \bar{\Lambda}_k } \Bigr]
\Bigr) \Bigr|_{\xi=0}
%\label{eq58}
\notag \\
= \sum \limits_{m} \, \left(m_1, \ldots m_n \right)! \, \prod \limits_{j=1}^n \Bigl( \kappa^{m_j
(j-1)} \Bigl[ L^{-1} \sum \limits_{k=0}^{L-1} \Psi(\bar{\Lambda}_k) \bar{\Lambda}_k^{m_j} \Bigr] \Bigr)
\label{eq59}.
\end{gather}
The left summation in \eqref{eq59} goes over all non-negative integers $m_i$ that satisfy the
condition $\sum_{i=0}^n m_i = l$, and $(m_1 \ldots m_n)! = (m_1+ \cdots +m_n)! / (m_1 ! \cdots m_n !)$
are the multinomial coefficient.
The inequality $|\bar{\Lambda}_k| =|\Lambda_k - \lambda_{\ast}| < 1$ implies $|\bar{\Lambda}_k^m| < 1$,
 $m \in \mbox{\textbf{N}}$. This allows us to obtain the upper estimates
\begin{equation*}
\Bigl| L^{-1} \sum \limits_{k=0}^{L-1} \Psi(\bar{\Lambda}_k) \bar{\Lambda}_k^{m} \Bigr| \le \max
\limits_{|\beta_k| < 1} \Bigl| L^{-1} \sum \limits_{k=0}^{L-1} \Psi(\bar{\Lambda}_k) \beta_k \Bigr| =
G_1 \le G,
%\label{eq60}
\end{equation*}
where $G=\max\{G_1, \, 1\}$. Notice that the number of multipliers  $ L^{-1} \sum_{k=0}^{L-1}
\Psi(\bar{\Lambda}_k) \bar{\Lambda}_k^{m}$ with $m \ne 0$ in \eqref{eq59} does not exceed the
least of the numbers $n$ and $l$. This allows us to write a sequence of upper estimates:
\begin{gather}
\Bigl| L^{-n} \sum \limits_{s} \Theta_n^{[s]} \left( \mathbf{s}_n, \, \mathbf{h}_n (\kappa)
\right)^l \Bigr|
\le \sum \limits_{m} \, \left(m_1, \ldots m_n \right)! \, \prod \limits_{j=1}^n
\Bigl( \kappa^{m_j (j-1)} \Bigl| L^{-1} \sum \limits_{k=0}^{L-1} \Psi(\bar{\Lambda}_k)
\bar{\Lambda}_k^{m_j} \Bigr| \Bigr)
\label{eq61} \\
\le G^l \, \sum \limits_{m} \, \left(m_1, \ldots m_n \right)! \, \prod \limits_{j=1}^n
\kappa^{m_j (j-1)} \le G^l (1-\kappa)^{-l}.
\label{eq62}
\end{gather}
As result, we see from \eqref{eq56}, \eqref{eq61} and \eqref{eq62} that
\begin{equation}
|\sigma_n| \le \sum \limits_{l=0}^{\infty} \frac{1}{l!} |f^{(l)} (\lambda_{\ast})| \,  G^l \le M \,
\sum \limits_{l=0}^{\infty} \frac{1}{l!} G^l = M\, e^G.
\label{eq63}
\end{equation}
Taking into account the relations \eqref{eq31}-\eqref{eq33},
\eqref{eq61}-\eqref{eq63}, we find:
\begin{gather}
|\sigma_{n+k} - \sigma_n| = \Bigl| \sum \limits_s L^{-n} \Theta_n^{[s]} \sum \limits_r L^{-k}
\Theta_k^{[r]} \left( f(\lambda _n^{[s]} + \Delta_{n,k}^{[r]})-f(\lambda _n^{[s]}) \right) \Bigr|
%\label{eq64}
\notag \\
\le \Bigl| \sum \limits_{l=1}^{\infty} \frac{1}{l!} \Bigl( \sum \limits_s L^{-n} \Theta_n^{[s]}
f^{(l)} (\lambda_n^{(s)}) \Bigr) \, \Bigl(\sum \limits_k L^{-k} \Theta_k^{[r]}  (\Delta_{n,k}^{(s)})^l
\Bigr) \Bigr|
%\label{eq65}
\notag \\
\le \sum \limits_{l=1}^{\infty} \frac{1}{l!} \Bigl| \sum \limits_s L^{-n} \Theta_n^{[s]} f^{(l)}
(\lambda_n^{(s)}) \Bigr| \, \kappa^{nl} (1-\kappa)^l \Bigl| \sum \limits_r L^{-k} \Theta_k^{[r]}
\left( \mathbf{r}_k, \, \mathbf{h}_k (\kappa) \right)^l \Bigr|
%\label{eq66}
\notag \\
\le M \, e^G \, \sum \limits_{l=1}^{\infty} \frac{1}{l!} G^l  \kappa^{nl} \le M e^G \,
( e^{G \kappa^n} -1 ).
%\label{eq67}
\notag
\end{gather}
We see that
\begin{equation*}
\forall \epsilon>0 \quad \exists N_{\epsilon} \quad \forall
n>N_{\epsilon}: \quad |\sigma_{n+k}-\sigma_{n+k'}| < 2 M e^G \, (
e^{G \kappa^n} -1 ) < \epsilon,
%\label{eq68}
\end{equation*}
which implies convergence of the Cauchy sequence $\{ \sigma_n \}$. 
$\blacksquare$

\section{Random processes transformations in a circuit with
dissipation and delayed feedback.}

Consider a circuit described by the equations \eqref{eq47},
\eqref{eq48} in assumption that $\tau_{\mbox{\scriptsize{ cor}}}
\sim T_{\mbox{\scriptsize{del}}}$ \cite{ZverRub1, ZverRub2, ZverRub3, Zverev2}. This
situation is more difficult for analysis as one should take into account
correlations between the variables $\xi_N^{\mbox{\scriptsize{fl}}} =
\xi ( t_0 + (N+1) T_{\mbox{\scriptsize{del}}} )$ with different
$N$. Now we are forced to seek the stationary solution of the
generalized Kolmogorov-Chapman equation for the multi-time
distribution functions:
\begin{gather}
P_{\mbox{\scriptsize{st}}} ((X_s),\xi_{n+1})= \int d \pmb{Y} d \pmb{\xi} \,
P_{\mbox{\scriptsize{st}}} ((Y_s),\xi_0)
\label{eq69} \\
\qquad \qquad \times \prod \limits_{p=0}^n \delta^{(2)} (X_p - \xi_p - F(Y_p)) \,
w( \xi_{p+1}, \xi_p , [\tau_{p+1} - \tau_p]).
%\label{eq70}
\notag
\end{gather}
In \eqref{eq69} we use a short-hand notation
\begin{equation*}
P_N((X_s),\xi) = P((X_0[NT_{\mbox{\scriptsize{del}}}], \,
X_1[NT_ {\mbox{\scriptsize{del}}} + \tau_1], \cdots ,
X_n[NT_{\mbox{\scriptsize{del}}} + \tau_n]), \, \xi[NT_{\mbox{\scriptsize{del}}}]),
%\label{eq71}
\end{equation*}
\noindent where $X_i, \, i=0,1,...,n$, are the signal amplitudes at
$NT_{\mbox{ \scriptsize{del}}}+\tau_i$, respectively, and $\xi$
denotes the amplitude of $\xi^{\mbox{\scriptsize{fl}}}$ at
$NT_{\mbox{\scriptsize{ del}}}$. We also require $0 = \tau_0 <
\tau_1 < \cdots < \tau_n < \tau_{n+1} = T_{\mbox{\scriptsize{del}}}$
and use a notation $d \pmb{A} \equiv dA_0 \, dA_1 \, dA_2 \ldots$.
The noise term $\xi^{\mbox{\scriptsize{fl}}}$ is assumed to be the
Markovian random process with arbitrary $\tau_{\mbox{\scriptsize{
cor}}}$ and the transition density function $w( \xi, \xi' , [\tau])$
(the Ornstein-Uhlenbeck and Kubo-Andersen random process were
treated in \cite{ZverRub2, ZverRub3, Zverev2}). The equation for the
distribution function Fourier transform reads
\begin{gather}
\psi_{\mbox{\scriptsize{st}}} ((U_s),\Omega_{n+1}) = \int d \pmb{V} d \pmb{\Omega} \,
\psi_{\mbox{\scriptsize{st}}}
((V_s), \Omega_0)
 \label{eq72} \\
\qquad \qquad \times \prod \limits_{p=0}^n \sigma (U_p, V_p) \,
H( \Omega_{p+1}, \Omega_p - U_p , [\tau_{p+1} - \tau_p]),
%\label{eq73}
\notag
\end{gather}
where $H(\Omega, \Omega', [\tau])$ is the Fourier transform of $w(
\xi, \xi' , [\tau])$ and the function $\sigma (U,V)$ is defined by
\eqref{eq5}. This function may be written in the form
\begin{equation*}
\sigma =\bar {\sigma} +\Delta \sigma , \quad \bar {\sigma}
\left( U,V \right) =\left( 2\pi \left| V\right| \right)^{-1}
e^{-i \, \mbox{\scriptsize{Re}}U}\delta \left( \left| V\right|-
\kappa \left| U\right| \right) ,
%\label{eq74}
\end{equation*}
\noindent where the first term $\bar{\sigma}$ corresponds to the
random phase approximation and describes the case of \textit{chaotic
motion} with \textit{intense phase mixing} \cite{ZverRub3, Zverev2}.
This means that in the limit $\lambda \to \infty$ in \eqref{eq48}
one can replace $\sigma$ by $\bar{\sigma}$ (respective approximate
solution of \eqref{eq69}, \eqref{eq72} are labeled by superscript
(0)). The exact and approximate forms of \eqref{eq69} can be
presented in the operator form:
\begin{equation}
\psi_{\mbox{\scriptsize{st}}}=\left( {\bf \hat K} + \varepsilon
{\bf \hat S}\right) \psi_{\mbox{\scriptsize{st}}} \quad \xrightarrow{\sigma
\to \bar{\sigma}, \quad \mbox{random phase approx.}} \quad
\psi_{\mbox{\scriptsize{st}}}^{\left( 0\right) }={\bf \hat K}
\psi _{\mbox{\scriptsize{st}}}^{\left( 0\right) }.
\label{eq75}
\end{equation}
\noindent In case of the Ornstein-Uhlenbeck random process
\cite{Zverev2} the explicit formula for ${\bf \hat K}$ takes the
form:
\begin{gather}
{\bf \hat K}f((U_s),U_{n+1}) = \Bigl[ \: \prod \limits_{p=0}^n e^{
- i \, \mbox{\scriptsize{Re}} U_p} \Bigr] \Phi ((U_s),U_{n+1})
\left\langle\!\!\!\left\langle f \Bigl( (\kappa U_s e^{i \phi_s}),
\sum \limits_{q=0}^{n+1} \theta_{\tau_q} U_q \Bigr)
\right\rangle\!\!\!\right\rangle _{\phi_s} ,
\label{eq76} \\
\Phi ((U_s),U_{n+1}) =\exp \Bigl\{ - \frac{R}{4} \, \sum \limits_{p=1}^{n+1}
{\Bigl| \sum \limits_{q=p}^{n+1} \frac{\theta_{\tau_q}}{\theta_{\tau_p}} U_q
\Bigr|^2} \Bigl( 1 - \frac{\theta_{\tau_p}^2}{\theta_{\tau_{p-1}}^2} \Bigr) \Bigr\}.
%\label{eq77}
\notag
\end{gather}
\noindent
It is evident that the operator ${\bf \hat K}$ produces a
\textit{scaling transformation (dilatation)}, because the right-hand side of 
\eqref{eq76} contains the functions with scaled arguments $\kappa U_s$. Thus 
the right equation in \eqref{eq75} is a \textit{generalized dilatation
equation}.  The equation \eqref{eq69} may be viewed as an equation with the 
"small" operator $\varepsilon {\bf \hat S}$. This enables to obtain the solution 
in the form of a series expansion. Taking into account the normalization conditions 
$\psi_{\mbox{ \scriptsize{st}}} \left( \left( 0 \right) ,0 \right) =
\psi_{\mbox{ \scriptsize{st}}}^{\left( 0 \right)} \left( \left( 0
\right) , 0 \right) =1$, one finds
\begin{equation*}
\psi _{\mbox{\scriptsize{st}}}^{\left( 0\right) }\left( \left( U_s\right) ,
\Omega \right) = \lim_{m\rightarrow \infty }{\bf \hat K}^m \, {\bf 1},
\qquad \qquad \qquad
\psi_{\mbox{\scriptsize{st}}}= \psi_{\mbox{\scriptsize{st}}}^{\left( 0 \right) }+
\sum \limits_{j=1}^\infty \varepsilon^j \Bigl\{ \Bigl( \sum \limits_{p=0}^\infty
{\bf \hat K}^p \Bigr) {\bf \hat S} \Bigr\}^j
\psi_{\mbox{\scriptsize{st}}}^{ \left( 0 \right) }.
%\label{eq78}
\end{equation*}
\noindent
This approach was developed in \cite{Zverev2} where the criterion of convergence was examined.

\subsection*{Acknowledgements}
VZ acknowledges the financial support from the European Mathematical
Society.


\begin{thebibliography}{99}

\bibitem{Zverev1} Zverev V.V., On the conditions for the existence of
fractal domain integrals, \textit{ Theor. Math. Phys.} \textbf{107}, (1996), 419-426.

\bibitem{Feder}  Feder J., Fractals, Plenum press, 1988.

\bibitem{Ikeda}  Ikeda K., Daido H., Akimoto O., Optical Turbulence:
chaotic behavior of transmitted light from a ring cavity, \textit{Phys. Rev. Lett.}
\textbf{45}, (1980), 709-712.

\bibitem{Singh}  Singh S., Agarwal G.S., Chaos in coherent
two-photon processes in a ring cavity, \textit{Opt. Comm.}
\textbf{47}, (1983), 73-76.

\bibitem{ZverRub1}  Zverev V.V., Rubinstein B.Y., Random self-modulation of radiation
in a ring cavity: Case of strong mixing \textit{Opt. Spect.} \textbf{65}, (1988), 971-978.

\bibitem{ZverRub2}  Zverev V.V., Rubinstein B.Y., Autostochasticity and
conversion of lasing fluctuations in a ring cavity with a nonlinear element,
\textit{Opt. Spect.} \textbf{70}, (1991), 1305-1311.


\bibitem{ZverRub3}  Zverev V.V., Rubinstein B.Y., Chaotic oscillations and noise
transformations in a simple dissipative system with delayed feedback,
\textit{J. Stat. Phys.} \textbf{63}, (1991), 221-239.

\bibitem{Zverev2}  Zverev V. V., Noise transformation in nonlinear system with intensity
dependent phase rotation, \textit{Stochastics and Dynamics.}
\textbf{3}, (2003), 421-433.



\end{thebibliography}
\end{document}